\documentclass[twocolumn,prl,amsmath,amssymb,floatfix,showpacs]{revtex4}

\usepackage{graphicx}% Include figure files
\usepackage{dcolumn}% Align table columns on decimal point
\usepackage{bm}% bold math

\newcommand{\ybro}{Y$_{2-x}$Bi$_x$Ru$_2$O$_7$}
\newcommand{\etal}{{\it et al.}}
\newcommand{\EF}{$E_{\rm F}$}

\begin{document}

\preprint{APS/123-QED}

\title{Core-Level X-Ray Photoemission Satellites in Ruthenates:\\
A New Mechanism Revealing the Mott Transition}

\author{Hyeong-Do Kim,$^1$ Han-Jin Noh,$^2$ K.-H. Kim,$^2$
and S.-J. Oh$^{2,*}$}
\affiliation{
$^1$Pohang Accelerator Laboratory, Pohang University of Science and Technology,
Pohang 790-784, Korea\\
$^2$School of Physics and Center for Strongly Correlated Materials
Research, Seoul National University, Seoul 151-742, Korea }

\date{\today}

\begin{abstract}
Ru 3$d$ core-level x-ray photoemission spectra of various ruthenates
are examined. They show in general two-peak structures, which can
be assigned as the screened and unscreened peaks. The screened
peak is absent in a Mott insulator, but develops into a main peak
in the metallic regime.  This spectral behavior is well explained
by the dynamical mean-field theory calculation for the single-band
Hubbard model with on-site core-hole potential using the exact
diagonalization method.  The new mechanism of the core-level
photoemission satellite can be utilized to reveal the Mott
transition phenomenon in various strongly correlated electron
systems, especially in nano-scale devices and phase-separated
materials.
\end{abstract}


\maketitle

%
% Introduction
%

Core-level x-ray photoemission spectroscopy (XPS) has long been a
powerful tool to investigate the chemical environment of solids
and molecules \cite{XPS}. Usually the variation of
the binding energy depending on the chemical environment
(``chemical shift'') is utilized for this purpose, but satellite
structures arising from the Coulomb attraction between valence
electrons and a core hole created in photoemission can be a
fruitful source of information, especially in strongly correlated
systems such as rare-earth \cite{GS} and transition-metal
compounds (TMCs) \cite{Zaanen86,Shen87}. A good example is the
high-$T_c$ and related cuprates \cite{Shen87}, where the satellite
structures of Cu 2$p$ core-level XPS have been instrumental to
confirm the charge-transfer nature of the insulating gap in the
parent compounds which were traditionally classified as Mott
insulators \cite{Mott90}, and determine basic parameters such as
the $d$-$d$ Coulomb correlation energy and the charge-transfer
energy.

\begin{figure}[t]
\includegraphics{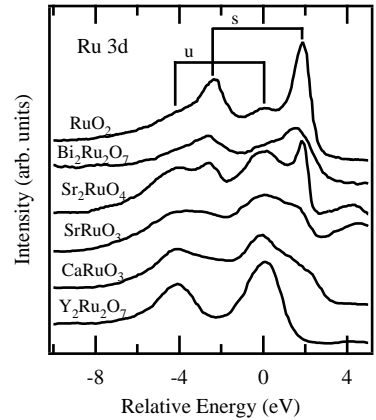}
\caption{Ru 3$d$ XPS spectra of RuO$_2$ (after \cite{Kim97}),
Bi$_2$Ru$_2$O$_7$ (after \cite{Cox83}), Sr$_2$RuO$_4$ 
(after \cite{Sekiyama00}), SrRuO$_3$, CaRuO$_3$, Y$_2$Ru$_2$O$_7$ 
(after \cite{Cox83}). ``s''
and ``u'' denote screened and unscreened peaks, respectively,
following \cite{Cox83}. All the spectra are aligned by the
unscreened-peak positions of Ru 3$d_{5/2}$. }
\end{figure}

Such a strong electron correlation effect has been considered to
be weak in 4$d$ TMCs because 4$d$ orbitals are fairly delocalized,
and RuO$_2$, for example, is traditionally classified as a band
metal \cite{Daniels84}. However, recent studies revealed that many
ruthenates such as Ca$_{2-x}$Sr$_x$RuO$_4$ \cite{Nakatsuji00} and
pyrochlores \cite{Kanno93,Lee97} have various interesting
properties related with the correlation effect among Ru 4$d$
electrons. Ru 3$d$ core-level XPS spectra also show some hint of
the strongly correlated nature of Ru 4$d$ electrons.  As shown in
Fig.~1, the Ru 3$d$ spin-orbit doublet, of which splitting is
about 4~eV, consists of roughly two components, ``screened''
(denoted as ``s'') and ``unscreened'' (denoted as ``u'') peaks in
various ruthenates \cite{Cox83,Kim97,Sekiyama00} including RuO$_2$
where their origin has been the subject of a long debate
\cite{Cox83,Kim97,Sekiyama00,Ruthenate,Okada02}. The peak ``s'',
located at about 2~eV lower binding energy than that of the rather
broad unscreened peak ``u'', is absent in an insulator
Y$_2$Ru$_2$O$_7$ and grows as the system becomes more metallic.
Hence Cox \etal\ suggested that the peak ``s'' might originate
from the screening of a core hole by a quasiparticle at \EF\
\cite{Cox83}. On the other hand, we may attribute it to the
charge-transfer mechanism as in 3$d$ TMCs, but the energy
separation between the main and satellite peaks is much smaller
than that of a Pd 3$d$ spectrum in PdO \cite{Uozumi00}, while we
expect larger separation according to the chemical trend
\cite{Zaanen86}. Quite recently, Okada \cite{Okada02} tried to
explain the two-peak structure in Sr$_2$RuO$_4$ \cite{Sekiyama00}
by the ``non-local screening'' effect \cite{Veenendaal93} in
addition to the charge-transfer mechanism. He could not, however,
fit the strong intensity variation and the little change of the
energy separation in ruthenate series with reasonable values of
parameters.

In this Letter, we propose a new mechanism for the two-peak
structures in the Ru 3$d$ core-level XPS spectra of ruthenates
based on a single-band Hubbard model.  The core-level spectra of a
single-band Hubbard model with core-hole potential are calculated
by the dynamical mean-field theory (DMFT) \cite{Georges96}.
The DMFT can describe successfully the Mott transition with the 
change of ratio of the Coulomb interacion $U$ between conduction 
electrons to the band width $W$, and it should be suitable for 
a local problem like core-level XPS despite its shortcomings due to
``local'' electron self-energy \cite{Inoue95}. In this picture,
the ``screened'' peak originates from the screening of a core-hole
by quasiparticles on the Fermi surface. We found that the
calculation can reproduce the spectral behavior shown in Fig.~1
just by changing the correlation strength with reasonable
parameter values.  We also made a systematic study of the Ru 3$d$
XPS spectra of a bandwidth-controlled Mott-Hubbard system \ybro\
\cite{Lee97} as a function of composition to strengthen our
conclusion.

%
% Calculational Method
%

We start from the single-band Hubbard model
\begin{eqnarray}
H&=&\sum_{k\sigma} \epsilon_k a^\dagger_{k\sigma} a_{k\sigma}
    + U \sum_i n_{i\uparrow} n_{i\downarrow},
\end{eqnarray}
where the notation is standard, which is exactly mapped in infinite dimensions
on to the single-impurity Anderson model according to the DMFT \cite{Georges96}
\begin{eqnarray}
H_{\rm And}&=&\sum_{l=2,\sigma}^{N_s} \tilde{\epsilon}_l
                c^\dagger_{l\sigma} c_{l\sigma}
              + \epsilon_d \sum_{\sigma} d^\dagger_\sigma d_\sigma
              + U n_{d\uparrow} n_{d\downarrow} \nonumber\\
              & &+ \sum_{l\sigma} V_l(c^\dagger_{l\sigma} d_\sigma
                 + {\rm H.c.}).
\end{eqnarray}
In order to describe a core-level spectrum, we should add a term
due to a core hole for XPS final states after getting the converged 
model parameters,
\begin{eqnarray}
H_{\rm core} = \bigl( \epsilon_h - Q \sum_{\sigma} d^\dagger_\sigma d_\sigma
               \bigr) h^\dagger h,
\end{eqnarray}
where the operator $h^\dagger$ ($h$) creates (destroys) a core
hole on the ``impurity" site, $\epsilon_h$ is a core-hole energy,
and $Q$ ($\equiv 1.25U$, for simplicity in all calculations) is the Coulomb
interaction between the core-hole and ``impurity-site" conduction electrons.
In order to solve the model with the semi-elliptical shape for
the conduction-band density of states, we followed the exact diagonalization
method proposed by Caffarel and Krauth \cite{Caffarel94}.

%
% Overall behavior
%

Figure 2 shows core-level and valence-band spectra of the ten-site
($N_s = 10$) half-filled model obtained by the modified Lanczos
method changing the ratio $W/U$ from a Mott insulator regime to a
good metal regime to show the behavior of spectral weights in the
bandwidth-controlled metal-insulator transition. For this
particle-hole symmetric model, $\epsilon_d$ is chosen to be
$-U/2$, then the chemical potential $\mu$ is identically zero. The
core-hole energy $\epsilon_h$ is set to $Q$. The core-level
spectra are mainly composed of two peaks (minor peaks are due to
the finite size) and their behavior with $W/U$ is quite similar to
that of the valence-band spectra and, moreover, the intensity of
the ``screened peak'' is found to be nearly proportional to the
quasiparticle weight $Z$ calculated by $( 1-\frac{\partial {\rm
Im} \Sigma(i\omega)}{\partial\omega})^{-1} \bigr|_{\omega
\rightarrow 0}$, which implies that the two-peak structure is
strongly related with the Mott transition. In a Mott insulator
($W/U = 0.65$) we can see a broad single peak, and as the coherent
peak grows with the increase of $W$ in the valence-band spectra, a
sharp shake-down satellite structure appears in the core-level
spectra in the positive energy side and eventually becomes a main
peak with the asymmetric low-energy tail.  Around $W/U = 2.0$
(top-most figure), both the core-level and valence-band spectra
remind of the famous 6~eV satellite in Ni metal, which is
explained by the Mahan-Nozi\`eres-DeDominicis model with strong
core-hole potential for the former and by the Hubbard model for
the latter \cite{Davis86}. Now, both spectra are explained in a
single model thanks to the DMFT based on the exact
diagonalization, and it should be applicable to other strongly
correlated systems showing the Mott transition.

\begin{figure}[t]
\includegraphics[scale=0.6]{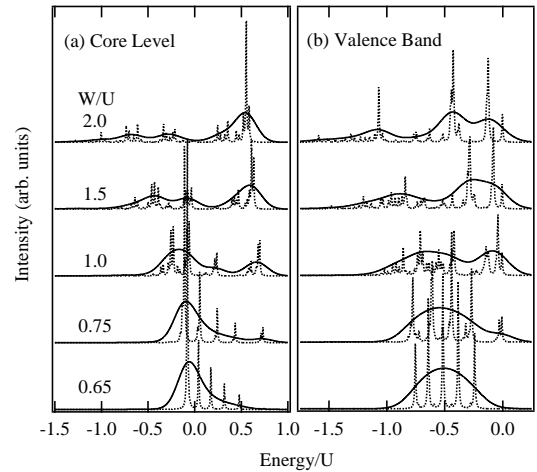}
\caption{(a) Core-level and (b) valence-band spectra of the
half-filled Hubbard model with $N_s = 10$ calculated by the DMFT 
varying $W/U$.
Solid lines are guidelines obtained by
broadening dotted lines with a Gaussian of 0.25 full width at half
maximum to remove discreteness due to the finite size. }
\end{figure}

%
% Explanation for the spectral behavior
%

\begin{figure}[b]
\includegraphics[scale=0.6]{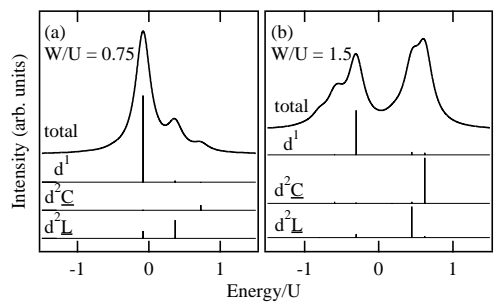}
\caption{Core-level spectra in the four-site model for (a) $W/U =
0.75$ and (b) 1.5. The calculated spectra are broadened by a
Lorentzian of 0.25 full width at half maximum.  The bar diagram at
the bottom indicates the weights of different valence-electron
configurations in each XPS final state. The three valence-electron
configurations  $d^1$, $d^2\underline{\rm C}$, and
$d^2\underline{\rm L}$ are shown schematically in Fig~3.}
\end{figure}

In order to investigate in detail the origin of the two-peak
structure in the core-level spectra, we obtained all the
eigenvalues and eigenvectors of the initial and final states of a
four-site model by explicit exact diagonalization.  Figure~3 shows
the calculated core-level spectra for the case of (a) $W/U = 0.75$
and (b) 1.5. Here we see again that for a narrow band case (a)
only the unscreened peak is dominant, while for the wide band (b)
both screened and unscreened peaks appear in the core-level
spectra.  To understand the valence-electron configuration of
these different core-level final states, we plot at the bottom of
the figure the bar diagram showing the weights of various
valence-electron configurations contributing to each core-level
peak.  For the four-site model, we should consider three energy
levels in the conduction band representing a lower Hubbard band
(LHB), a coherent peak, and an upper Hubbard band, along with the
``impurity" level at the core-hole site in the presence of a
core-hole.   Then there are three major valence-electron
configurations in the XPS final state, i.e. $d^1$ ($\equiv d^1
{\rm L^2 C^1}$, for $N_s = 4$), $d^2\underline{\rm C}$, and
$d^2\underline{\rm L}$, where $d$ denotes the ``impurity" level
and C (L) denote the coherent peak (LHB). These valence-electron
configurations are shown schematically in Fig.~4. It can be easily
noticed from the bar diagram at the bottom of Fig.~3 that the
unscreened peak originates from the configuration $d^1$, while the
screened peak from the configurations $d^2\underline{\rm C}$ and
$d^2\underline{\rm L}$.

With these configurations, the behavior of two-peak structure with
the change of $W/U$ is simply understood.  When $W/U$ is so small
that the system becomes a Mott insulator, the configuration
$d^2\underline{\rm C}$ is absent both in the ground and the final
states, so the unscreened peak, of which main configuration is
$d^1$, becomes dominant, and the screened peak by the
configuration $d^2\underline{\rm L}$ is located near the
unscreened peak (the separation is $Q-U$ in the limit $W/U$
$\rightarrow$ 0). Since the configuration $d^2\underline{\rm L}$
represents a single-hole state in the LHB, the core-level peak of
a Mott insulator has extra linewidth broadening reflecting the LHB
shape. When $W/U$ is close to the critical value and the system
becomes a metal, the configuration $d^2\underline{\rm C}$ is now
available to take part in the screening of the core hole, which
makes a sharp shake-down satellite peak, whose energy is about
$Q-U/2$ lower than that of the unscreened peak and its intensity
increases with the weight of the coherent peak.

\begin{figure}[t]
\includegraphics[scale=0.35]{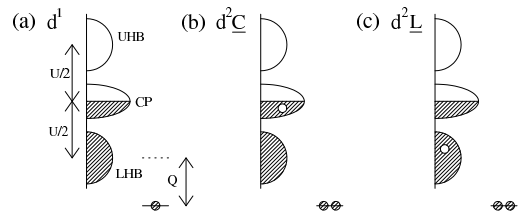}
\caption{Schematic diagrams of three major valence-electron
configurations contributing to the core-level spectra.  Note that
the ``impurity" level at the core-hole site is lowered by the
attractive Coulomb potential $Q$. }
\end{figure}

%
% Comparison with Experimental Results
%

To test the validity of the present model, we made a systematic
study of the Ru 3$d$ core-levels in \ybro\ system.  This system is
known to be a bandwidth-controlled Mott-Hubbard system from
theoretical calculation \cite{Lee97}, transport properties
\cite{Kanno93} and valence-band photoelectron spectra
\cite{Park01}, which shows the metal-insulator transition around
$x = 0.9$ \cite{Kanno93}. We took Ru 3$d$ XPS spectra of
polycrystalline \ybro\ samples ($x = 0.0$, 0.4, 1.0, 1.6, and 2.0)
using Mg K$\alpha$ source after annealing four hours under the oxygen
partial pressure of 10$^{-5}$~Torr at 900~K.  The annealing
process, which removes almost all of the carbon contaminations in
the sample, is essential because the binding energy of the Ru 3$d$
core-level is close to that of C 1$s$ peak.

\begin{figure}[b]
\includegraphics[scale=0.6]{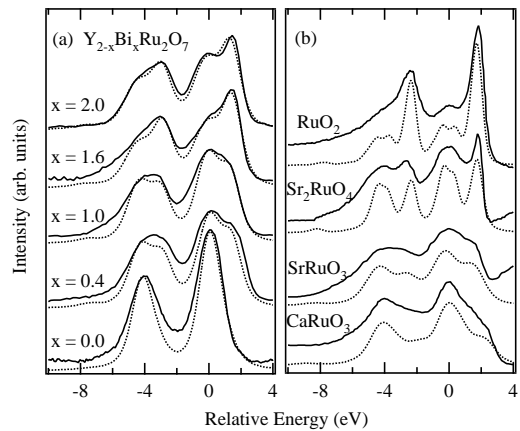}
\caption{Comparison of Ru 3$d$ XPS spectra (solid lines) of (a)
\ybro\ and (b) other ruthenates  with the present model
calculations (dotted lines). See text for details. }
\end{figure}

Ru $3d$ XPS spectra (solid lines) of \ybro\ thus obtained are
shown in Fig.~5 (a) after removing the inelastic background and the
small residual C 1$s$ peak contribution by deconvolution using
Doniach-$\check{\rm S}$unsi\'c line shapes. We can see the
systematic behavior that the weight of the screened peak becomes
larger as the Bi concentration increases. This tendency is
strongly correlated with the transport properties \cite{Kanno93}
and valence-band photoelectron spectra \cite{Park01}. Using the
present model, the core-level spectra of \ybro\ series could be
reproduced by changing only $W$ with fixed $U$ value ($=1.7$~eV). We
also fit the spectra of other ruthenates shown in Fig.~1 by
freeing $U$, and Table~1 summarizes all the parameter values.
After proper broadening of calculation results by a Voigt function
to simulate the core-hole lifetime and the experimental
resolution, final results (dotted lines) are plotted over the
experimental spectra in Fig.~5 (a) \ybro\ and (b) other
ruthenates. The fitting results are quite satisfactory and all the
spectral behaviors are well reproduced. Rather surprisingly,
RuO$_2$, which was believed to be well described by band
calculations \cite{Daniels84}, should be reconsidered as a
strongly correlated metal according to our fitting result ($W/U =
2.0$).

\begin{table}[b]
\caption{Parameter values (in eV) obtained from fitting Ru 3$d$
XPS spectra of ruthenates. }
\begin{ruledtabular}
\begin{tabular}{cccccc}
$x$ in \ybro\ & $U$ & $W$ &     & $U$ & $W$ \\
\hline
0.0 & 1.7 & 1.2 &   CaRuO$_3$   & 2.7 & 2.6 \\
0.4 & 1.7 & 2.1 &   SrRuO$_3$   & 2.15& 2.6 \\
1.0 & 1.7 & 2.15& Sr$_2$RuO$_4$ & 2.15& 2.8 \\
1.6 & 1.7 & 2.9 &    RuO$_2$    & 1.8 & 3.6 \\
2.0 & 1.7 & 2.7 &               &     &     \\
\end{tabular}
\end{ruledtabular}
\end{table}

Although the present model does not include such terms as
4$d$-orbital degeneracy and also the Ru 4$d_{t_{2g}}$ band shape
is far from semi-elliptical \cite{Ishii00,Bands}, it is quite
successful in describing the two-peak structure of the core-level
spectra, which confirms that the Mott-Hubbard mechanism is the
most important factor.  Nevertheless, the role of O 2$p$ bands
could not be ruled out in two respects: (1) The change of
$W$-value for \ybro\ is rather larger than expected
\cite{Lee97,Ishii00}, while $W$ values for other ruthenates are
comparable to those of band calculations \cite{Bands}. Since we
are dealing with an {\it effective} Hubbard model, in which O 2$p$
bands are integrated out, the $U$ value should change in real
materials depending on the Ru 4$d$-O 2$p$ hybridization strength,
which would result in the reduction of the $W$ variation. (2) We
may expect that a similar two-peak structure should appear in
other strongly correlated TMCs. Actually, there is some hint of it
in Ti$_2$O$_3$ \cite{Uozumi96} and V$_2$O$_3$ \cite{Park94}. But
the screened-peak intensity is rather small in these systems
compared with the coherent peak of the valence band \cite{Mo03},
which may be explained by the competition between the
charge-transfer screening by the O 2$p$ electron and the screening
by quasiparticles at $E_{\rm F}$. Since the quasiparticle
screening takes place through mediating O 2$p$ orbitals, it is
probably more rapidly suppressed than the charge-transfer
screening as the hybridization strength becomes weak. Further
experimental and theoretical studies are necessary to clarify this
issue.

%
% Summary
%

In summary, we have elucidated the origin of two-peak structures
in the Ru 3$d$ core-level XPS spectra of various ruthenates 
as due to the different screening mechanisms in the
Mott-Hubbard picture of the metal-insulator transition. This
mechanism should be applicable to other TMCs
as well, hence core-level XPS can be utilized to distinguish
metallic from insulating regions in phase-separated materials
\cite{Mathur03} and TMC-based nano-electronic devices 
when used in conjuction with
spectro-microscopic tools such as scanning photoemission
microscopy.
%
% Acknowledgments
%

\begin{acknowledgments}
This work was supported in part by the Korea Science and
Engineering Foundation through the Center for Strongly Correlated
Materials Research of Seoul National University and by the Korea
Ministry of Science and Technology.
\end{acknowledgments}

%
% References
%


\begin{thebibliography}{99}

\bibitem[*]{}
Corresponding author.

\bibitem{XPS}
T. A. Carlson, {\it Photoelectron and Auger Spectroscopy} (Plenum Press,
New York, 1975);
M. Cardona and L. Ley in {\it Photoemission in Solids I} (Springer-Verlag,
Berlin, 1978).

\bibitem{GS}
O. Gunnarsson and K. Sch\"onhammer, Phys. Rev. Lett. {\bf 50}, 604 (1983);
J. W. Allen \etal, Adv. Phys. {\bf 35}, 275 (1986).

\bibitem{Zaanen86}
J. Zaanen, C. Westra, and G. A. Sawatzky, Phys. Rev. B {\bf 33}, 8060 (1986);
J. Park \etal, Phys. Rev. B {\bf 37}, 10867 (1988);
G. Lee and S.-J. Oh, Phys. Rev. B {\bf 43}, 14674 (1991);
A. E. Bocquet \etal, Phys. Rev. B {\bf 46}, 3771 (1992).

\bibitem{Shen87}
Z.-X. Shen \etal, Phys. Rev. B {\bf 36}, 8414 (1987).

\bibitem{Mott90}
N. F. Mott, {\it Metal-Insulator Transitions} (Taylor and Francis, New York,
1990).

\bibitem{Daniels84}
R. R. Daniels \etal, Phys. Rev. B {\bf 29}, 1813 (1984).

% \bibitem{Zaanen85}
% J. Zaanen, G. A. Sawatzky, and J. W. Allen, Phys. Rev. Lett. {\bf 55}, 418
% (1985).

\bibitem{Nakatsuji00}
S. Nakatsuji and Y. Maeno, Phys. Rev. Lett. {\bf 84}, 2666 (2000).

\bibitem{Kanno93}
R. Kanno \etal, J. Solid State Chem. {\bf 102}, 106 (1993);
S. Yoshii and M. Sato, J. Phys. Soc. Jpn. {\bf 68}, 3034 (1999).

\bibitem{Lee97}
K.-S. Lee, D.-K. Seo, and M.-H. Whangbo, J. Solid State Chem. {\bf 131},
405 (1997).

\bibitem{Kim97}
Y. J. Kim, Y. Gao, and S. A. Chambers, Appl. Surf. Sci. {\bf 120}, 250 (1997).

\bibitem{Cox83}
P. A. Cox \etal, J. Phys. C: Solid State Phys. {\bf 16}, 6221 (1983).

\bibitem{Sekiyama00}
A. Sekiyama, {\it Abstracts of the Meeting of the Phys. Soc. Jpn}
(55th Annual Meeting, Niigata, Sep. 2000); recited from \cite{Okada02}.

\bibitem{Ruthenate}
K. S. Kim and N. Winograd, J. Catal. {\bf 35}, 66 (1974);
H. J. Lewerenz, S. Stucki, and R. Kotz, Surf. Sci. {\bf 126}, 893 (1983);
K. Reuter and M. Scheffler, Surf. Sci. {\bf 490}, 20 (2001);
H. Over \etal, Surf. Sci. {\bf 504}, L196 (2002).

\bibitem{Okada02}
K. Okada, Surf. Rev. Lett. {\bf 9}, 1023 (2002).

\bibitem{Uozumi00}
T. Uozumi \etal, J. Phys. Soc. Jpn. {\bf 69}, 1226 (2000).

\bibitem{Veenendaal93}
M. A. van Veenendaal and G. A. Sawatzky, Phys. Rev. Lett. {\bf 70}, 2459 (1993).

\bibitem{Georges96}
For a review, see A. Georges \etal, Rev. Mod. Phys. {\bf 68}, 13 (1996).

\bibitem{Inoue95}
I. H. Inoue \etal, Phys. Rev. Lett. {\bf 74}, 2539 (1995).

\bibitem{Caffarel94}
M. Caffarel and W. Krauth, Phys. Rev. Lett. {\bf 72}, 1545 (1994).

\bibitem{Davis86}
For a review, see L. C. Davis, J. Appl. Phys. {\bf 59}, R25 (1986).

% \bibitem{V2O3}
% The V 2$p$ XPS spectrum of metallic V$_2$O$_3$ [J.-H. Park, Ph.D. thesis,
% University of Michigan, 1994] shows a small shake-down peak absent
% in an insulating phase,
% but its intensity is too small compared with the coherent-peak intensity
% in the valence-band spectrum [S.-K. Mo \etal, Phys. Rev. Lett. {\bf 90},
% 186403 (2003)],
% which may be due to strong charge transfer from O 2$p$ to V 3$d$
% orbitals, evident from a high-energy satellite structure,
% suppressing the screening of a core hole by quasiparticles on the Fermi
% surface.

% \bibitem{Okada91}
% K. Okada and A. Kotani, J. Phys. Soc. Jpn. {\bf 60}, 772 (1991).

\bibitem{Park01}
J. Park, Ph.D thesis, Seoul National University, 2001.

\bibitem{Ishii00}
F. Ishii and T. Oguchi, J. Phys. Soc. Jpn. {\bf 69}, 526 (2000).

\bibitem{Bands}
K. M. Glassford and J. R. Chelikowsky, Phys. Rev. B {\bf 47}, 1732 (1993);
T. Oguchi, Phys. Rev. B {\bf 51}, 1385 (1995);
I. I. Mazin and D. J. Singh, Phys. Rev. B {\bf 56}, 2556 (1997).

\bibitem{Uozumi96}
T. Uozumi \etal, J. Phys. Soc. Jpn. {\bf 65}, 1150 (1996).

\bibitem{Park94}
J.-H. Park, Ph.D. thesis, University of Michigan, 1994.

\bibitem{Mo03}
S.-K. Mo \etal, Phys. Rev. Lett. {\bf 90}, 186403 (2003).

\bibitem{Mathur03}
N. Mathur and P. Littlewood, Phys. Today {\bf 56}, 25 (2003).


\end{thebibliography}
\end{document}